# Long-range self-hybridized exciton polaritons in two-dimensional Ruddlesden-Popper perovskites

Maximilian Black[1,*], Mehdi Asadi[2], Parsa Darman[2], Finja Schillmöller[1], Sara Darbari[2,*], Nahid Talebi[1,*]

[1] Institute of Experimental and Applied Physics, Kiel University, 24098 Kiel, Germany
[2] Nano-Sensors and Detectors Lab., Faculty of Electrical and Computer Engineering, Tarbiat Modares University, Tehran, Iran

**ABSTRACT:** Lead halide perovskites have emerged as platforms for exciton-polaritonic studies at room temperature thanks to their excellent photoluminescence efficiency and synthetic versatility. In this work we find proof of strong exciton-photon coupling in cavities formed by the layered crystals themselves, a phenomenon known as self-hybridization effect. We use multi-layers of high quality Ruddlesden-Popper perovskites in their 2D crystalline form, benefitting from their quantum-well excitonic resonances and the strong Fabry-Perot cavity modes resulting from the total-internal-reflection at their smooth surfaces. Optical spectroscopy reveals bending of the cavity modes typical for exciton-polariton formation, and photoluminescence spectroscopy shows thickness dependent splitting of the excitonic resonance. Strikingly, local optical excitation with energy below the excitonic resonance of the flakes in photoluminescence measurements unveils coupling of light to in-plane polaritonic modes with directed propagation. These exciton-polaritons exhibit high coupling efficiencies and extremely low loss propagation mechanisms which is confirmed by finite difference time domain simulations. We therefore prove that mesoscopic 2D Ruddlesden-Popper perovskites flakes represent an effective but simple system to study the rich physics of exciton-polaritons at room temperature.

**Introduction**

Strong light-matter interactions are central to many physical and chemical phenomena, ranging from coherent control of chemical reactions to the formation of new states of matter. Being first described as superpositions of light and matter excitations[1,2], polaritons form when an elementary material excitation interacts strongly with an electromagnetic radiation field. Most notably excitons in a semiconductor can strongly interact with confined light such as the resonant modes of a cavity and the resulting quasi-particle is hence called exciton-polariton (E-P). E-Ps exhibit exceptional properties such as low effective mass, long coherence lengths and enhanced nonlinearities[3], opening up possibilities to study effective photon-photon interactions leading to Bose-Einstein condensation[4,5] (BEC) and many-body states in the strong-coupling regime[6,7]. However, it is challenging to transfer the strong nonlinearities and subsequently polariton condensation to room temperature. Therefore, it is necessary to use the control parameter of E-P density as substitute for the temperature, expressing the need for high exciton yield semiconductors with sufficiently strong exciton binding energies. To realize E-Ps, a variety of semiconductors have been combined with different ways to confine light and to achieve overlaps with excitonic wavefunctions. Prominently, transition-metal dichalcogenides (TMDCs) have been used due to their exceptionally high exciton binding energies of a few hundred eV and high oscillator strengths for effective light-matter coupling particularly in the case of monolayers[8,9]. In practice, TMDC monolayers have been merged with high-quality microcavities[10], photonic and plasmonic crystals[11], plasmonic nanoantennas[12] and plasmon polaritons[13]. Yet handling of TMDC monolayers proves to be quite challenging and they barely reach lateral dimensions exceeding several tens of micrometers.

Lead-halide perovskites have proven to be a suitable alternative due to their remarkably easy synthesis, direct bandgaps, high optical gain and high oscillator strength[14,15]. In particular, of special interest are quasi-two-dimensional Ruddlesden-Popper organo-metal halide perovskite (RPP) layers that have been applied in solar cells in 2015[16] and have been used for realizations of photodetectors, LEDs and lasers[17–19]. These layers are quantum-well-structured materials of a certain numbers of perovskite layers (well) which are separated by spacer molecules (barrier) along the $z$-axis[20–22]. For these quasi-two-dimensional materials, exceptionally high exciton binding energies of several hundred meV have been reported in bulk crystals[23–26] and connected to quantum confinement decreasing the likelihood of electron-hole pairs dissociating at charge separation interfaces, which additionally results in a high emission quantum yield. For perovskites polariton formation has been realized with a variety of photonic structures[27] such as plasmonic lattices[28] and microcavities[29–31] with superfluidity claimed to be reached in the latter systems[32]. While high quality optical cavities incorporating Debye-Bragg reflectors provide a high tunability of the cavity resonances to the excitonic energies, their complex fabrication schemes hinder their implementation into versatile optoelectronic devices. In addition, plasmons provide an evanescent electromagnetic field close to metal-dielectric interfaces that can strongly interact with excitons especially when localized plasmons or plasmonic lattices with tunable collective lattice resonances are involved[33]. Nevertheless, plasmons sustain significant dissipative losses which necessitate precise nanofabrication processes. Therefore, there is a fundamental need for simple and efficient platforms that enable robust exciton-photon coupling.

Recently, waveguide modes of thin semiconductor crystals strongly interacting with excitons have gained interest, a process called self-hybridization. Tens of nanometers thin TMDC layers have been studied by optical microspectroscopy[34,35] and scanning near-field optical microscopy[36], where both exciton-photon anticrossing behavior and E-P propagation have been detected[37]. Additionally, using cathodoluminescence spectroscopy, where the confined Cherenkov radiation strongly couples to the excitons[38], propagation mechanisms of E-Ps with high spatial resolution, edge E-Ps, as well as polariton-polariton interactions were explored[39]. Furthermore, it has been shown that the energy band gap of plasmonic crystals can be tuned by combining the thin TMDC layers with plasmonic crystals due to the strong coupling of exciton polaritons with the plasmonic Bloch modes[40,41].

While strong luminescence from excitons has been observed only for mono- and bilayer TMDCs, excitons in bulk RPP crystals retain their two-dimensional property of high binding energies and prominent luminescence peaks in the bulk crystal. Large single-crystal flakes of quasi-two-dimensional perovskites have been shown to feature polariton-related bending of optical cavity modes in their reflection spectra[42] and spin-dependent exciton-exciton interactions[43]. Furthermore, energy transfer from higher to lower energy E-Ps has been shown in RPPs on gold substrates by direct laser excitation of lower polariton branches and ultrafast spectroscopy[44,45]. However, a direct proof of propagating exciton polaritons in freestanding thin RPP crystals and an extensive investigation of their propagation properties and length have yet to be reported.

In this work we investigate the optical properties of thin exfoliated RPP flakes by means of photoluminescence (PL), reflection and absorption spectroscopy combined with analytical and finite element calculations of the layered system. We find a clear anti-crossing of cavity modes in the vicinity of the exciton energy in reflection and absorption spectra in an excellent agreement with our theoretical results. By applying site-selective PL excitation of lower polariton branches and coupling out of E-Ps at the edges of our flakes we confirm long-range propagation of polaritons across the waveguide. These findings are confirmed with finite-difference time-domain simulations predicting extremely low loss propagating waves at the measured energies. Our results prove that even simple quasi-confined modes of a two-dimensional RPP waveguide coupled to excitons yield high quality E-Ps at room temperature. This will spark further investigations utilizing the intriguing fabrication simplicity toward BEC in these straightforward systems.

**Results**

We mechanically exfoliate sub-micrometer thin flakes of two-dimensional Ruddlesden-Popper perovskites (2D RPP) from macroscopic sheets of bulk crystals onto glass substrate. The bulk RPP $(BA)_2PbI_4$ is grown in a solution and its chemical structure is schematically depicted in Figure 1a, showing its layered configuration of monolayers of the metal-halide $PbI_4^{2-}$ which are separated by butylammonium cations ($BA^+$). The monolayers themselves consist of octahedra formed by lead (II) cations surrounded by six iodide anions. We note that we use single atomic layers (n=1) of these octahedra, which implies the absence of anions that stabilize the octahedra found in the crystal structure of multiple layer (n>1) 2D RPPs or 3D perovskites[27]. The monolayers host excitonic excitations which exhibit quantum confinement as a result of being sandwiched by organic cations, consequently resulting in high binding energies of the excitons. Additionally, the structure of such stacked quantum wells enables the mechanical exfoliation of the bulk crystal to sub-micrometer thin flakes. Moreover, the dielectric constant of the barrier ($BA^+$) is less than that of the inorganic well ($PbI_4^{2-}$) which contributes to high exciton binding energy according to dielectric confinement phenomena. Aliphatic molecules such as BA have been shown to have greater exciton binding energies compared with aromatic molecules such as PEA, which makes $(BA)_2PbI_4$ an interesting candidate for studying exciton-polaritons (E-Ps) within two dimensional Ruddlesden-Popper perovskite flakes[46]. We showcase the formation of waveguide modes within exfoliated thin RPP flakes in Figure 1a schematically. Total internal reflection from the boundaries causes the thin crystal to act as a cavity itself with different cavity modes and with their optical dispersion being dependent on the thickness of the flake. Due to the wave function overlap of the excitons and the confined optical modes they hybridize to polaritonic waveguide modes of which the electric field distribution of the ground mode has been simulated and is depicted in Figure 1a.

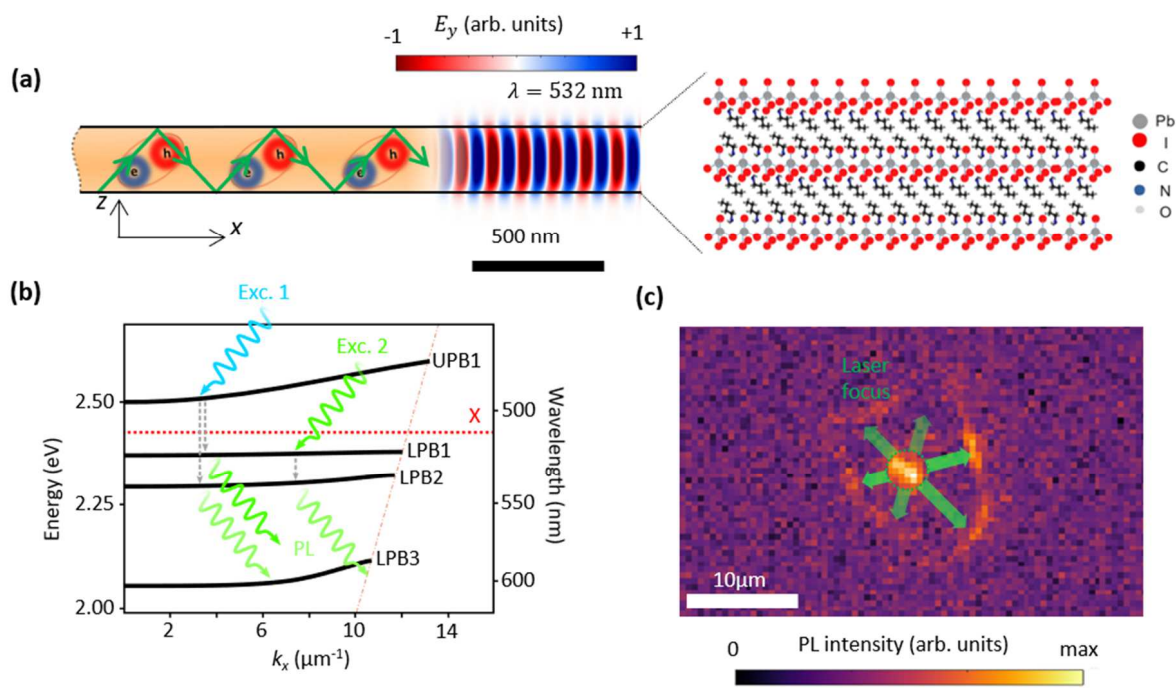

Figure 1. (a) Self-hybridized exciton-polaritons in thin RPP crystals. Excitons supported by the stacked quantum-well structure of the RPP depicted on the right side interact with the waveguide modes that stem from reflection at the boundaries of the thin film, leading to polariton formation. As an example, the y-component of the electric field of the resulting first-order even mode is shown. (b) Excitation and radiation schemes of the measured photoluminescence (PL) signal. Light with energy lower or higher than the exciton resonance couples to lower or upper polariton branches (LPB and UPB), respectively. The excited states undergo relaxation processes to the LPBs situated below via either non-radiative or radiative paths, where the emission within the infrared to the visible range is detected (green wiggly arrows pointing to the bottom-right). The LPB coupling to light causes a detectable PL signal. (c) PL intensity map of an RPP flake excited by a 532 nm CW laser focused at its centre. Excited lower polariton modes propagate to the edges of the flake where they couple to light.

An exemplary dispersion relation of these E-Ps is given in Figure 1b showing the characteristic avoided crossing of the cavity modes with the excitonic resonance, where the dispersion of the latter can be approximated to be constant in low-momentum ranges. Here, the presence of the excitons and the subsequent hybridization causes a bending of the photonic modes away from the excitonic resonance dividing their dispersion into upper and lower polariton branches (UPB and LPB, respectively). To investigate the energy transfer between different polariton branches we employ two excitation schemes indicated in Figure 1b. First, by choosing the energy range of the incident light we selectively excite the first UPB, which transfers its energy to LPBs through radiative decay, where possible radiative emission during this process falls within long wavelengths beyond the detection limit of our detectors. When the LPBs couple directly to light, a detectable photoluminescence (PL) signal is emitted. Second, we excite the highest LPB directly by illuminating the RPP flake with a CW laser with a wavelength of 532 nm just above the excitonic resonance of 510.8 nm which then relaxes to LPBs at lower energies. We note the mere presence of a detectable PL signal in Figure 1c indicates the occurrence of this process. Strikingly, the PL intensity map shows PL signal not only at the focal point of the laser in the center of the flake but also at the edges, indicating E-Ps being excited at the laser spot and propagating as waveguide modes towards the edge where they can couple to light again. Thus, we report direct observation of long-range propagation of self-hybridized E-P in thin RPP flakes.

To justify this conclusion in detail, we investigate the formation of polaritons using reflection, transmission and absorption spectroscopy. The exfoliated flakes show a broad variety of colors dependent on their thickness which is easily visible in the reflection image in Figure 2a. This directly stems from the Fabry-Pérot resonances occurring in the meso-scale films, causing wavelength-dependent reflection and transmission peaks. To explore the momentum-dependent reflection, transmission and absorption spectra of the sample, we utilize a model system composed of a layer of RPP with a known thickness surrounded by vacuum and solve the wave equation for this system versus the photon energy and the incidence angle. To introduce the nonlinearity of the RPP into this classical calculation we use measured data of the dielectric function of $(BA)_2PbI_4$ for the RPP layer[47]. We particularly observe an excellent agreement between the s-polarized incident light in our calculations compared to measured results, as expected, due to the mainly in-plane polarization of the light in our measurement schemes.

Since the microscope objective used to measure the spectra in Figure 2b has a numerical aperture of 0.7 and thus a collection angle of 0° to 44.43° we average our calculations over this angle range and compare the resulting spectra to the measured spectra in Figure 2b. For given values of thickness we observe a remarkable agreement with the measurements confirming the accuracy of our calculations and allowing conclusions to be drawn about the dispersion relations. Both in the reflection and the transmission energy-momentum maps in Figures 2c and 2d several cavity modes are evident, showing a clear avoidance of the excitonic resonance and subsequently a bending of their dispersion. Additionally, in all of the reflection spectra and in the reflection energy-momentum maps, a peak coincides with the exciton resonance. While from a standalone absorber anomalous dispersion is to be expected, the appearance of a reflection peak at the exciton resonance is characteristic of the absorber interacting with the cavity modes[35]. Thus we observe LPBs below the exciton energy as dips and peaks in the transmission and reflection spectra respectively with varying degree of hybridization of the exciton with the cavity modes. Moreover, even the spectral signatures of an UPB above the exciton resonance are observed which are present in both calculated and measured spectra.

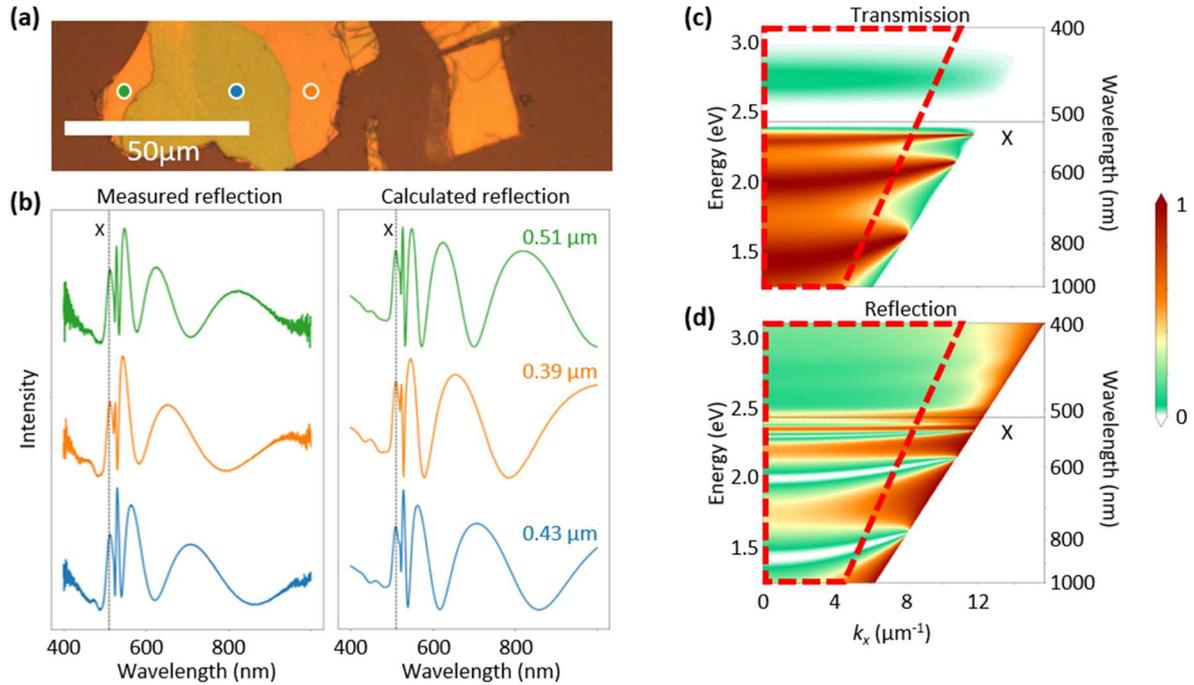

Figure 2. (a) Real-space reflection image of the investigated RPP flake with color-coded positions of the measured reflection spectra. (b) Measured (left) and calculated (right) reflection spectra of the marked positions. The calculations take into account the collection angle range 0° to 44.43° of the microscope objective (NA=0.7) and are therefore an average over the corresponding in-plane wave numbers $k_x = (2\pi/\lambda)\sin(\theta)$ of the energy-momentum maps, where $\lambda$ is the wavelength and $\theta$ is the incidence angle. The marked positions are at visibly different heights of the flake and agree perfectly well with the calculated spectra for selected thicknesses. (c) Calculated energy versus momentum transmission and (d) reflection maps for a thin film with the thickness of 430 nm. The red trapezoids mark the area corresponding to the average over the incidence angle range. The fabry-perot resonances bend due to strong coupling with the excitons at the energy marked by the dotted line. Formation of lower and upper polariton branches below and above the exciton energy are observed both in the reflection and the transmission map.

To further solidify the conclusion of the polaritonic nature of the observed modes, we acquire the PL spectra of the same positions marked in Figure 2a. The PL spectra in Figure 3a were taken using a band-pass filter with 472 nm center wavelength and 30 nm width in the excitation path and subsequently a 520 nm center wavelength and 35 nm width band-pass filter in the emission path, thus we directly excite in the region of the UPBs. In all three positions a broad PL signal is observed that is red-shifted with respect to the excitonic resonance, consisting of a superposition of multiple peaks with varying positions depending on the flake thickness. The most intense signal is observed in the orange-colored spectrum, having the highest energy and the lowest thickness dependence. However, the second peak appears closer to the first and has a similar intensity. We identify the peaks as different LPBs with the first LPB exhibiting the strongest coupling of the cavity mode to the exciton, leading to a pronounced bending of its dispersion. This yields the flattest dispersion and the least sensitivity to thickness for this mode, while the energy of lower-energy LPBs experience a larger dependence on the thickness of the flake. Indicated by the varying intensity, the energy transfer to the LPBs is dependent on the acquisition position. The most efficient transfer occurs from the excited UPB to the first LPB while the relaxation to the second LPB becomes more effective at specific thicknesses of the flakes. We note that the excitation scheme allows both the relaxation from the UPB and the first LPB to the second LPB, both contributing to its elevated intensity in the orange spectrum.

The latter process of energy transfer between LPBs is better configured by the PL signal in Figure 3c, where the flake marked in Figure 3b is excited with a 532 nm CW laser focused onto its left edge. Since the exciton at 510.8 nm cannot be excited in this way, the observed signal has to stem from an LPB. The PL spectra in Figure 3d are taken at the opposing edges of the flake and show the tail of an LPB cut off by the combination of dielectric mirror and edge-pass emission filter with its measured transmittance depicted in Figure 3e. Similar to Figure 1, the excited mode emits light at the position of the laser focus and at the opposing edge. This means that depending on their in-plane momentum component, the exciton polaritons are guided through the waveguide formed by the RPP flake itself and couple to light at its edge. Additionally, as the edge PL signal is strongest in the direction perpendicular to the edge where the laser spot couples light into the flake, we deduce that the propagation direction is dependent on the shape of the coupling edge. Since the in-plane momentum is expected to determine the efficiency of modes being guided along the flake, the light coupled out at the opposing edge should, on average, be emitted at a higher angle compared to the incidence angle. Therefore, we expect the signal at the opposing edge to be slightly blue-shifted, thus causing the detected tail of the LPB to be of weaker intensity. Remarkably, even considering this factor, the lower spectrum in Figure 3d representing the waveguided PL signal at that edge is not notably weaker than that at the position of the laser spot. We thus conclude that we observe long-range propagating self-hybridized E-Ps with a controllable propagation direction in thin RPP flakes.

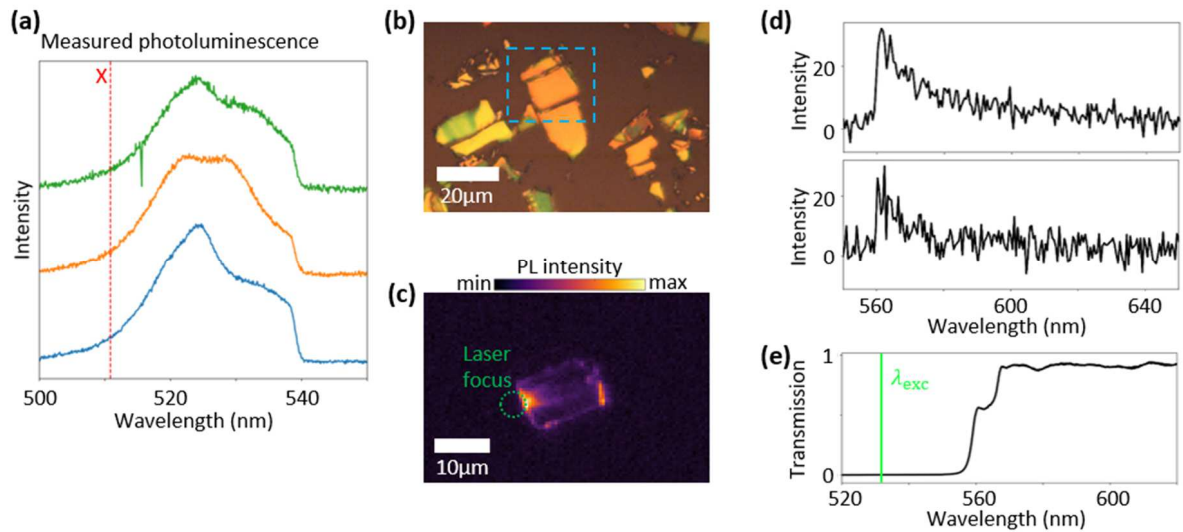

Figure 3. (a) PL spectra of the color-coded positions in figure 2b, acquired using a 472 nm $\pm$ 15 nm excitation and a 525 nm $\pm$ 25 nm emission filter. The spectra show multiple peaks redshifted with respect to the exciton energy with thickness dependent peak positions. (b) Reflection image of the RPP flakes used for the laser induced PL measurements in (c) and (d). (c) PL intensity maps with laser excitation at the edge of the RPP flakes. A clear PL signal from the edge positioned opposite to the laser focus (green circle) demonstrates the directional propagation of polaritonic waves along the flakes which couple to the light at the edges. (d) The PL spectra corresponding to (c) at the left (upper) and the right edge (lower) with laser excitation at the left edge of the flake. The sharp drop just below 560 nm corresponds to the lowest wavelength that is transmitted by the long-pass filter and dichroic mirror combination used as emission filter. With laser excitation at 532 nm, the excitons at 511 nm are not excited, but the laser couples directly to an LPB. (e) Transmittance spectrum of the emission filter used for the laser-based PL setup. The spectra in (d) have been corrected by the shape of this transmittance.

Having established the formation of E-Ps in our thin RPP slab waveguides we briefly address here their dependence on the thickness of the flakes. Hence, we plot the spectra calculated in the same fashion as in Figure 2c and 2d versus the flake thickness in Figure 4. The resulting thickness-dependent dispersions show polaritonic features already discussed, such as the reflection peak at the exciton resonance. Moreover, they exhibit polaritonic behavior at every elevated thickness. We see both UPBs and LPBs most clearly in the reflection plot of Figure 4a and the absorption plot in Figure 4c. Their bending is enhanced in the vicinity of the exciton energy which agrees with the lowered thickness sensitivity of the first LPBs in the PL spectra. Additionally, a permanent dip in the absorption versus thickness plot resides at the exciton energy linked to the polaritonic Rabi splitting.

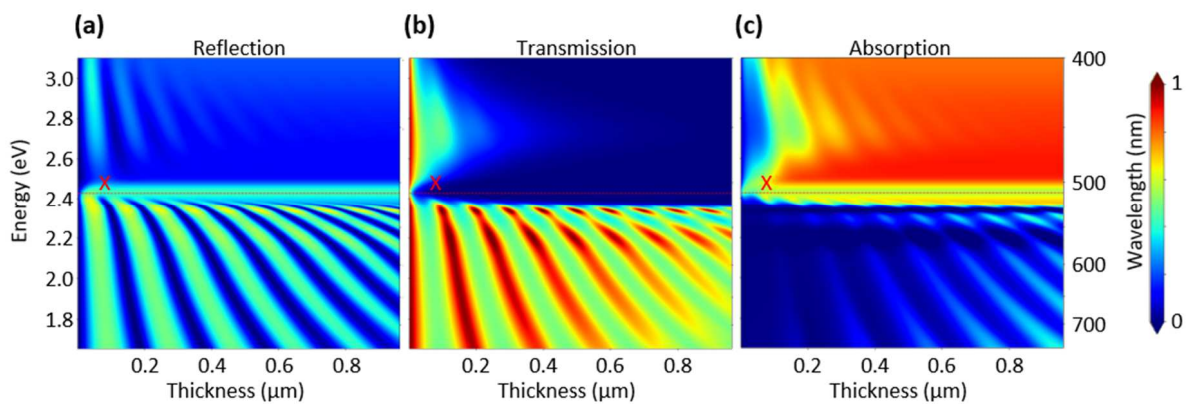

Figure 4. Calculated (a) reflection, (b) transmission and (c) absorption spectra of thin RPP crystals that are plotted versus the photon energy and crystal thickness. Corresponding to the calculated and measured spectra in Figure 2a, 2b and 2d, the spectra were averaged over the angle range 0° to 44.43°. With increasing thickness multiple fabry-perot resonances appear that show significant bending in the vicinity of the exciton energy marked by the grey line. Remarkably, the absorption spectra show a dip at the exciton energy due to the Rabi splitting in the strong-coupling regime.

Complementary to the measured propagation mechanisms of E-Ps, we investigate the coupling efficiency to E-Ps at the edge of a thin film and the propagation behavior of E-Ps along the RPP flake theoretically. Here, a monochromatic TE-polarized Gaussian beam is illuminated on the left edge of a 2D RPP film with the 430 nm thickness, with an incidence angle of 40° with respect to the horizontal axis. We show cross sections of the normalized electric field distribution in Figure 5a. While Figure 5a reveals no coupling at the incident wavelength of 500 nm, a large coupling efficiency into the semi-infinite 2D RPP flake is evident at 640 nm. Moreover, we find a multi-mode excitation showing nearly lossless propagation along the *x*-direction of the infinite flake with the latter owing to the negligible imaginary component of the permittivity at this energy range. With the same parameters and illumination geometry as in the latter case but considering a flake with a finite length of 15 µm, the confined light again efficiently couples out to the radiation continuum at the right edge. This simulation result is in accordance with the observed experimental result in Figure 3, pointing out the long-range propagation of the LPBs and their coupling to light at the opposing edge. The fine interference fringes in the bottom part of Figure 5a are explained by the partial reflection of the guided waves from the right edge.

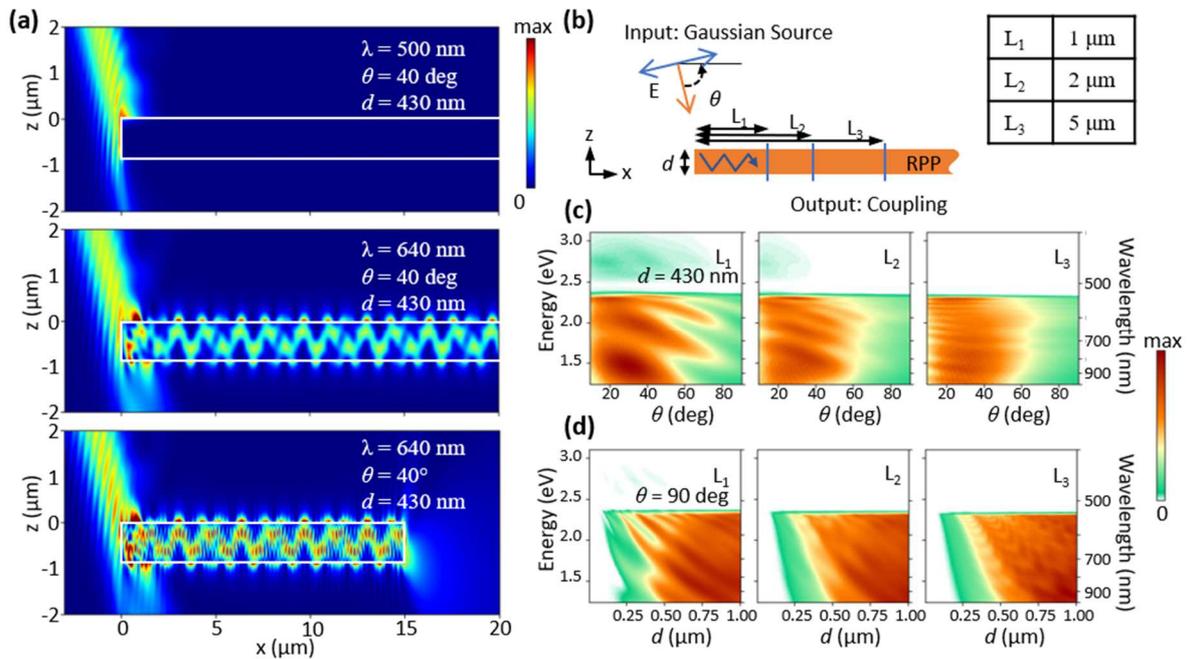

Figure 5: Numerical simulation of the coupling of incident light to the 2D RPP waveguide and subsequent propagation. (a) Cross sections of the electric field distribution of a monochromatic TE-polarized Gaussian beam illuminating a 430 nm thick flake at an angle of $\theta = 40°$ with respect to the horizontal axis. The incident beam is illuminating the left edge of an infinite flake at respective wavelengths of 500 nm (top) and 640 nm (center) and of a finite flake with the length of 15 µm at the wavelength of 640 nm (bottom). (b) Schematic cross-section view of the illumination geometry with flake thickness $d$ and incidence angle $\theta$. Three monitors $L_1$, $L_2$ and $L_3$ are probing the output power at different distances from the left edge, exploring the optical coupling of the beam to waveguide modes and their propagation along the flake infinite in *x*-direction: (c,d) Energy-resolved ratio of the output power at the different monitors to the incident input power plotted versus (c) incidence angle $\theta$ and (d) thickness $d$.

Furthermore, we investigate the coupling efficiencies and the consequent propagating behavior in a semi-infinite flake by placing three vertical output monitors $L_1$, $L_2$ and $L_3$ at different distances from the left edge, as indicated by the blue lines in the schematic in Figure 5b. The energy-resolved transmission values at the monitors, achieved by dividing the output optical power by the incident beam power, are plotted versus the incident angle in Figure 5c and versus the flake thickness in Figure 5d.

Figure 5c reveals that the optical coupling into the flake occurs for almost all the incident energies lower than the exciton energy at angles up to about $60°$, which can be attributed to the fact that scattering at the edge closes the momentum gap to all possible waveguide modes at nearly all possible in-plane momentum values. Moreover, the optical coupling is observed to be maximized within the angle range of $\theta = 20°$ to $40°$ for different energies, while the coupling strength is diminished for incident angles larger than about $40°$. As the coupled wave propagates along the flake we observe close to no diminishing of the mean power between $L_1$ and $L_3$, the only difference being interference patterns in $L_1$ due to the propagating wave overlapping with the incident beam. A similar picture is drawn by Figure 5d. Again, we observe very low loss between $L_1$ and $L_3$. It is evident that as the flake thickness increases the coupling efficiency increases as well resulting from the emergence of numerous modes that the incident optical beam can be coupled into. While very thin flakes host no cavity modes, the coupling efficiency raises from $d = 0.25$ μm at the energy ranges just below the excitonic resonance since Fabry-Pérot modes appear at low wavelengths first. This theoretical investigation confirms nearly perfect optical coupling of the incident beam to the E-Ps in the flake at the edges and a nearly lossless propagation at long distances, reinforcing our finding of long-range self-hybridized E-Ps.

**Conclusion**

In summary, we have probed the formation and long-range propagation of self-hybridized exciton polaritons in sub-micrometer thin two-dimensional Ruddlesden-Popper perovskite waveguides by means of transmission, reflection, and photoluminescence spectroscopy. Our experimental results are further supported by theoretical dispersion calculations and exploring the coupling efficiency of the optical waves into the exciton polaritons. The slab waveguide formed by the thin Ruddlesden-Popper perovskite crystal yields strong coupling between cavity modes and excitons, and thus the ensuing polariton formation modulates the optical response of the material, which is observable in the optical dispersion. This enables the absorption of photons both below and above the bandgap into lower and upper polariton branches respectively, as well as energy transfer between these branches. We directly probe these processes by selecting the exciting photon energy and detecting sub-bandgap emission. Due to the dielectric nature of the Ruddlesden-Popper perovskite below the excitonic energy, the lower polariton branches exhibit highly effective energy transmission via the guided modes across the thin flake with minimal indications of loss. By localized illumination with a sub-exciton-energy laser we show the direct excitation of these modes and their relaxation into lower energy polariton branches, causing photoluminescence signal at the edges far away from the laser spot. This demonstrates that light can couple to these guided modes either directly or through scattering at the edges, closing a possible in-plane momentum mismatch between incident light and propagating polariton mode (e.g. at normal incidence). Consequently, this opens up a wide range of excitation schemes, determining the propagation direction not only by the incident light but also by defined structures in the vicinity of the Ruddlesden-Popper perovskite layers. Furthermore, the polariton dispersion is tuned by the flake thickness, contributing to the parameter space to control the polariton propagation. Thus, our results pave new ways to investigate exciton polaritons in a system with intriguing simplicity and a high degree of versatility, encouraging future research into the characteristics of exciton-polaritons and their associated effective photon-photon interactions.

**Materials and methods**

*Sample preparation*

For the synthesis of the 2D RPP bulk crystal, first n-butylammonium iodide (BAI) was synthesized in a separate setting by slowly adding 25 ml 57% w/w aqueous hydriodic acid (HI) to 5 ml n-butylamine (BA) in an ice bath stirring for four hours. Second, 500 mg of PbO powder was dissolved in a mixture of HI solution (57%, 3 ml), and $H_3PO_2$ solution (50%, 850 μl), while heating at 120°C and stirring for about 5 minutes, which led to a bright yellow solution product of $PbI_2$. Then, 1.5 ml of the synthesized n-$CH_3(CH_2)_3NH_3I$ (BAI) was added to the $PbI_2$ solution, initially producing a black precipitate, which was subsequently dissolved by continued heating at 120°C. Stirring was stopped after 5 minutes, and the solution was left to cool down to room temperature, during which orange rectangular-shaped crystalline flakes are formed. Then, using vacuum filtering, $BA_2PbI_4$ orange rectangular flakes were separated from the solvent and left to dry for a few days. After that, the flakes were deposited on a substrate by mechanical exfoliation where a flake was picked up by a piece of scotch tape and after folding and unfolding for several times several flakes with reduced thicknesses to hundreds of nanometers were achieved which then were stamped on the substrate.

*Optical spectroscopy*

Reflectance and transmittance were recorded from the RPP flakes exfoliated on glass using a Nikon Eclipse Ti2-A inverted microscope and its built-in halogen lamps, in an assembled system provided by New Technologies and Consulting Company. Both were normalized by dividing through the reflectance of a silver mirror and the transmittance of glass respectively. The photoluminescence measurements were carried out with a halogen lamp, filtered by a 472 nm ± 15 nm excitation filter, and with a 532 nm CW laser. The low power of the halogen lamp within the excitation band proved to be sufficient for preventing photobleaching of the perovskites and for the same reason the laser power was damped by neutral density filters. To separate the excitation illumination, which was incident from the optical reflection path, from the emission signal, dichroic mirrors were employed in combination with bandpass and edgepass filters. The optical spectra were measured in a Princeton Instruments SpectraPro HRS 500-S grating spectrometer, using a reflective grating with 150 grooves/mm for the reflection/absorption while for photoluminescence measurements, a grating with 600 grooves/mm was utilized. All measurements were performed at room temperature.

*Analytical and numerical calculations*

For the analytical dispersion calculation, a layered model was defined with infinite extension in x- and y-direction, which consists of vacuum, a material with relative permittivity $\varepsilon_r(\omega)$ of RPP, and then vacuum again, all arranged along the z-direction in the sequential order. With light incident from the positive z-direction, the propagation is calculated in x-direction and the permittivity is assumed to be isotropic, which yields:

$$k_z = \pm k_0 \sqrt{\epsilon_r(\omega) - \sin^2(\theta)}.$$

Here, $k_z$ is the in-plane wavenumber of the propagating wave, $k_0$ denotes the wavenumber of the incoming radiation and $\theta$ its angle of incidence. Defining $z_1 = d$ and $z_2 = -d$ as the positions of the boundaries, the wave equation of for s-polarized incident light yielding the time-harmonic solution of the wave equation, becomes:

$$E_y = \begin{cases} E_0 e^{ik_x x} \left( e^{-ik_z^{(0)}(z-d)} + R_s e^{ik_z^{(0)}(z-d)} \right) & \text{if } z > d \\ E_0 e^{ik_x x} \left( C_1 e^{-ik_z(z-d)} + C_2 e^{ik_z(z+d)} \right) & \text{if } d > z > -d \\ T_s E_0 e^{ik_x x - ik_z^{(0)}(z+d)} & \text{if } -d > z \end{cases}.$$

with reflection amplitude $R_S$, transmission amplitude $T_S$ and the constant amplitudes $C_1$ and $C_2$. A similar expression for $H_x$ is found by $\vec{H} = \frac{1}{i\omega\mu_0} \vec{\nabla} \times \vec{E} = \frac{1}{i\omega\mu_0} \left( \frac{\partial E_y}{\partial x} \hat{z} - \frac{\partial E_y}{\partial z} \hat{x} \right)$ and applying the boundary conditions that $E_y$ and $\frac{\partial E_y}{\partial x}$ have to be continuous at $d = z$ and $d = -z$ results in the following equation system:

$$1 + R_s = C_1 + C_2 e^{2ik_z d}$$
$$ik_0 \cos(\theta)(-1 + R_s) = ik_z(-C_1 + C_2 e^{2ik_z d})$$
$$T_s = C_1 e^{2ik_z d} + C_2$$
$$-ik_0 \cos(\theta) T_s = ik_z(-C_1 e^{2ik_z d} + C_2)$$

The solutions for $R_S$ and $T_S$ represent the reflectance $|R_S|^2$ and transmittance $|T_S|^2$. It should be noted that a plane wave focused by the microscope objective can interfere with itself, hence taking the averages of $R_S$ and $T_S$ is done before taking their absolute values when calculating the averaged reflectance and transmittance over the corresponding range of incident angles.

In the case of the simulations of the coupling and propagation efficiency the finite difference time domain (FDTD) was utilized as the numerical method. Perfectly matched layers (PML) were used as the boundary conditions at the boundaries of the simulated box for which Maxwell's equations have been solved. Here, a TE-polarized beam was incident at the edge of an RPP layer with semi-infinite or finite extension in *x*-direction and infinite extension in *y*-direction and the resulting electric field distribution was calculated.

**Acknowledgement**

This project has received funding from the European Research Council (ERC) under the European Union's Horizon 2020 research and innovation programme under grant agreement no. 802130 (Kiel, NanoBeam), grant agreement no. 101157312 (Kiel, UltraCoherentCL) and grant agreement no. 101017720 (eBEAM), and from Deutsche Forschungsgemeinschaft under grant agreement no. 447330010 and Initiation of International Collaboration (TA 1694/6-1).

**References**

(1) Pekar, S. I. Theory of Electromagnetic Waves in a Crystal with Excitons. *J. Phys. Chem. Solids* **1958**, *5* (1), 11–22. https://doi.org/10.1016/0022-3697(58)90127-6.

(2) Hopfield, J. J. Theory of the Contribution of Excitons to the Complex Dielectric Constant of Crystals. *Phys. Rev.* **1958**, *112* (5), 1555–1567. https://doi.org/10.1103/PhysRev.112.1555.

(3) Carusotto, I.; Ciuti, C. Quantum Fluids of Light. *Rev. Mod. Phys.* **2013**, *85* (1), 299–366. https://doi.org/10.1103/RevModPhys.85.299.

(4) Kasprzak, J.; Richard, M.; Kundermann, S.; Baas, A.; Jeambrun, P.; Keeling, J. M. J.; Marchetti, F. M.; Szymańska, M. H.; André, R.; Staehli, J. L.; Savona, V.; Littlewood, P. B.; Deveaud, B.; Dang, L. S. Bose–Einstein Condensation of Exciton Polaritons. *Nature* **2006**, *443* (7110), 409–414. https://doi.org/10.1038/nature05131.

(5) Deng, H.; Haug, H.; Yamamoto, Y. Exciton-Polariton Bose-Einstein Condensation. *Rev. Mod. Phys.* **2010**, *82* (2), 1489–1537. https://doi.org/10.1103/RevModPhys.82.1489.

(6) Ma, R.; Saxberg, B.; Owens, C.; Leung, N.; Lu, Y.; Simon, J.; Schuster, D. I. A Dissipatively Stabilized Mott Insulator of Photons. *Nature* **2019**, *566* (7742), 51–57. https://doi.org/10.1038/s41586-019-0897-9.

(7) Carusotto, I.; Houck, A. A.; Kollár, A. J.; Roushan, P.; Schuster, D. I.; Simon, J. Photonic Materials in Circuit Quantum Electrodynamics. *Nat. Phys.* **2020**, *16* (3), 268–279. https://doi.org/10.1038/s41567-020-0815-y.

(8) Mak, K. F.; Shan, J. Photonics and Optoelectronics of 2D Semiconductor Transition Metal Dichalcogenides. *Nat. Photonics* **2016**, *10* (4), 216–226. https://doi.org/10.1038/nphoton.2015.282.

(9) Selig, M.; Berghäuser, G.; Raja, A.; Nagler, P.; Schüller, C.; Heinz, T. F.; Korn, T.; Chernikov, A.; Malic, E.; Knorr, A. Excitonic Linewidth and Coherence Lifetime in Monolayer Transition Metal Dichalcogenides. *Nat. Commun.* **2016**, *7* (1), 13279. https://doi.org/10.1038/ncomms13279.

(10) Flatten, L. C.; He, Z.; Coles, D. M.; Trichet, A. A. P.; Powell, A. W.; Taylor, R. A.; Warner, J. H.; Smith, J. M. Room-Temperature Exciton-Polaritons with Two-Dimensional WS2. *Sci. Rep.* **2016**, *6* (1), 33134. https://doi.org/10.1038/srep33134.

(11) Liu, W.; Lee, B.; Naylor, C. H.; Ee, H.-S.; Park, J.; Johnson, A. T. C.; Agarwal, R. Strong Exciton–Plasmon Coupling in MoS2 Coupled with Plasmonic Lattice. *Nano Lett.* **2016**, *16* (2), 1262–1269. https://doi.org/10.1021/acs.nanolett.5b04588.

(12) Stührenberg, M.; Munkhbat, B.; Baranov, D. G.; Cuadra, J.; Yankovich, A. B.; Antosiewicz, T. J.; Olsson, E.; Shegai, T. Strong Light–Matter Coupling between Plasmons in Individual Gold Bi-Pyramids and Excitons in Mono- and Multilayer WSe2. *Nano Lett.* **2018**, *18* (9), 5938–5945. https://doi.org/10.1021/acs.nanolett.8b02652.

(13) Sang, Y.; Wang, C.-Y.; Raja, S. S.; Cheng, C.-W.; Huang, C.-T.; Chen, C.-A.; Zhang, X.-Q.; Ahn, H.; Shih, C.-K.; Lee, Y.-H.; Shi, J.; Gwo, S. Tuning of Two-Dimensional Plasmon–Exciton Coupling in Full Parameter Space: A Polaritonic Non-Hermitian System. *Nano Lett.* **2021**, *21* (6), 2596–2602. https://doi.org/10.1021/acs.nanolett.1c00198.

(14) Smith, M. D.; Connor, B. A.; Karunadasa, H. I. Tuning the Luminescence of Layered Halide Perovskites. *Chem. Rev.* **2019**, *119* (5), 3104–3139. https://doi.org/10.1021/acs.chemrev.8b00477.

(15) La-Placa, M.-G.; Longo, G.; Babaei, A.; Martínez-Sarti, L.; Sessolo, M.; Bolink, H. J. Photoluminescence Quantum Yield Exceeding 80% in Low Dimensional Perovskite Thin-Films via Passivation Control. *Chem. Commun.* **2017**, *53* (62), 8707–8710. https://doi.org/10.1039/C7CC04149G.

(16) Cao, D. H.; Stoumpos, C. C.; Farha, O. K.; Hupp, J. T.; Kanatzidis, M. G. 2D Homologous Perovskites as Light-Absorbing Materials for Solar Cell Applications. *J. Am. Chem. Soc.* **2015**, *137* (24), 7843–7850. https://doi.org/10.1021/jacs.5b03796.

(17) Liang, Y.; Shang, Q.; Wei, Q.; Zhao, L.; Liu, Z.; Shi, J.; Zhong, Y.; Chen, J.; Gao, Y.; Li, M.; Liu, X.; Xing, G.; Zhang, Q. Lasing from Mechanically Exfoliated 2D Homologous Ruddlesden–Popper Perovskite Engineered by Inorganic Layer Thickness. *Adv. Mater.* **2019**, *31* (39), 1903030. https://doi.org/10.1002/adma.201903030.

(18) Chen, K.; Zhang, Q.; Liang, Y.; Song, J.; Li, C.; Chen, S.; Li, F.; Zhang, Q. Quasi-Two Dimensional Ruddlesden-Popper Halide Perovskites for Laser Applications. *Front. Phys.* **2023**, *19* (2), 23502. https://doi.org/10.1007/s11467-023-1347-6.

(19) Gao, X.; Zhang, X.; Yin, W.; Wang, H.; Hu, Y.; Zhang, Q.; Shi, Z.; Colvin, V. L.; Yu, W. W.; Zhang, Y. Ruddlesden–Popper Perovskites: Synthesis and Optical Properties for Optoelectronic Applications. *Adv. Sci.* **2019**, *6* (22), 1900941. https://doi.org/10.1002/advs.201900941.

(20) Dehghani, Z.; Nadafan, M.; Mohammadzadeh Shamloo, M. B.; Shadrokh, Z.; Gholipour, S.; Rajabi Manshadi, M. H.; Darbari, S.; Abdi, Y. Investigation of Dielectric, Linear, and Nonlinear Optical Properties of Synthesized 2D Ruddlesden-Popper-Type Halide Perovskite. *Opt. Laser Technol.* **2022**, *155*, 108352. https://doi.org/10.1016/j.optlastec.2022.108352.

(21) Darman, P.; Yaghoobi, A.; Darbari, S. Pinhole-Free 2D Ruddlesden–Popper Perovskite Layer with Close Packed Large Crystalline Grains, Suitable for Optoelectronic Applications. *Sci. Rep.* **2023**, *13* (1), 8374. https://doi.org/10.1038/s41598-023-35546-1.

(22) Mohammadzadeh Shamloo, M. B.; Darman, P.; Darbari, S.; Abdi, Y. Highly Stable and Sensitive Broadband Photodetector Based on BA2MAPb2I7/Si Heterojunction. *Opt. Laser Technol.* **2024**, *176*, 110889. https://doi.org/10.1016/j.optlastec.2024.110889.

(23) Ishihara, T.; Takahashi, J.; Goto, T. Exciton State in Two-Dimensional Perovskite Semiconductor (C10H21NH3)2PbI4. *Solid State Commun* **1989**, *69*, 933.

(24) Ishihara, T.; Takahashi, J.; Goto, T. Optical Properties Due to Electronic Transitions in Two-Dimensional Semiconductors (CnH2n+1NH3)2PbI4. *Phys Rev B Condens Matter Mater Phys* **1990**, *42*, 11099.

(25) Hong, X.; Ishihara, T.; Nurmikko, A. V. Dielectric Confinement Effect on Excitons in PbI4-Based Layered Semiconductors. *Phys Rev B Condens Matter Mater Phys* **1992**, *45*, 6961.

(26) Ishihara, T.; Hong, X.; Ding, J.; Nurmikko, A. V. Dielectric Confinement Effect for Exciton and Biexciton States in PbI4-Based Two-Dimensional Semiconductor Structures. *Surf Sci* **1992**, *267*, 323.

(27) Su, R.; Fieramosca, A.; Zhang, Q.; Nguyen, H. S.; Deleporte, E.; Chen, Z.; Sanvitto, D.; Liew, T. C. H.; Xiong, Q. Perovskite Semiconductors for Room-Temperature Exciton-Polaritonics. *Nat. Mater.* **2021**, *20* (10), 1315–1324. https://doi.org/10.1038/s41563-021-01035-x.

(28) Park, J.-E.; López-Arteaga, R.; Sample, A. D.; Cherqui, C. R.; Spanopoulos, I.; Guan, J.; Kanatzidis, M. G.; Schatz, G. C.; Weiss, E. A.; Odom, T. W. Polariton Dynamics in Two-Dimensional Ruddlesden–Popper Perovskites Strongly Coupled with Plasmonic Lattices. *ACS Nano* **2022**, *16* (3), 3917–3925. https://doi.org/10.1021/acsnano.1c09296.

(29) Su, R.; Wang, J.; Zhao, J.; Xing, J.; Zhao, W.; Diederichs, C.; Liew, T. C. H.; Xiong, Q. Room Temperature Long-Range Coherent Exciton Polariton Condensate Flow in Lead Halide Perovskites. *Sci. Adv.* **2018**, *4* (10), eaau0244. https://doi.org/10.1126/sciadv.aau0244.

(30) Spencer, M. S.; Fu, Y.; Schlaus, A. P.; Hwang, D.; Dai, Y.; Smith, M. D.; Gamelin, D. R.; Zhu, X.-Y. Spin-Orbit–Coupled Exciton-Polariton Condensates in Lead Halide Perovskites. *Sci. Adv.* **2021**, *7* (49), eabj7667. https://doi.org/10.1126/sciadv.abj7667.

(31) Li, Y.; Ma, X.; Zhai, X.; Gao, M.; Dai, H.; Schumacher, S.; Gao, T. Manipulating Polariton Condensates by Rashba-Dresselhaus Coupling at Room Temperature. *Nat. Commun.* **2022**, *13* (1), 3785. https://doi.org/10.1038/s41467-022-31529-4.

(32) Peng, K.; Tao, R.; Haeberlé, L.; Li, Q.; Jin, D.; Fleming, G. R.; Kéna-Cohen, S.; Zhang, X.; Bao, W. Room-Temperature Polariton Quantum Fluids in Halide Perovskites. *Nat. Commun.* **2022**, *13* (1), 7388. https://doi.org/10.1038/s41467-022-34987-y.

(33) Törmä, P.; Barnes, W. L. Strong Coupling between Surface Plasmon Polaritons and Emitters: A Review. *Rep. Prog. Phys.* **2014**, *78* (1), 013901. https://doi.org/10.1088/0034-4885/78/1/013901.

(34) Wang, Q.; Sun, L.; Zhang, B.; Chen, C.; Shen, X.; Lu, W. Direct Observation of Strong Light-Exciton Coupling in Thin WS2 Flakes. *Opt. Express* **2016**, *24* (7), 7151–7157. https://doi.org/10.1364/OE.24.007151.

(35) Munkhbat, B.; Baranov, D. G.; Stührenberg, M.; Wersäll, M.; Bisht, A.; Shegai, T. Self-Hybridized Exciton-Polaritons in Multilayers of Transition Metal Dichalcogenides for Efficient Light Absorption. *ACS Photonics* **2019**, *6* (1), 139–147. https://doi.org/10.1021/acsphotonics.8b01194.

(36) Hu, F.; Luan, Y.; Scott, M. E.; Yan, J.; Mandrus, D. G.; Xu, X.; Fei, Z. Imaging Exciton–Polariton Transport in MoSe2 Waveguides. *Nat. Photonics* **2017**, *11* (6), 356–360. https://doi.org/10.1038/nphoton.2017.65.

(37) Mrejen, M.; Yadgarov, L.; Levanon, A.; Suchowski, H. Transient Exciton-Polariton Dynamics in WSe2 by Ultrafast near-Field Imaging. *Sci. Adv.* 5 (2), eaat9618. https://doi.org/10.1126/sciadv.aat9618.

(38) Chahshouri, F.; Taleb, M.; Diekmann, F. K.; Rossnagel, K.; Talebi, N. Interaction of Excitons with Cherenkov Radiation in WSe2 beyond the Non-Recoil Approximation. *J. Phys. Appl. Phys.* **2022**, *55* (14), 145101. https://doi.org/10.1088/1361-6463/ac453a.

(39) Taleb, M.; Davoodi, F.; Diekmann, F. K.; Rossnagel, K.; Talebi, N. Charting the Exciton–Polariton Landscape of WSe2 Thin Flakes by Cathodoluminescence Spectroscopy. *Adv. Photonics Res.* **2022**, *3* (1), 2100124. https://doi.org/10.1002/adpr.202100124.

(40) Davoodi, F.; Talebi, N. Plasmon–Exciton Interactions in Nanometer-Thick Gold-WSe2 Multilayer Structures: Implications for Photodetectors, Sensors, and Light-Emitting Devices. *ACS Appl. Nano Mater.* **2021**, *4* (6), 6067–6074. https://doi.org/10.1021/acsanm.1c00889.

(41) Davoodi, F.; Taleb, M.; Diekmann, F. K.; Coenen, T.; Rossnagel, K.; Talebi, N. Tailoring the Band Structure of Plexcitonic Crystals by Strong Coupling. *ACS Photonics* **2022**, *9* (7), 2473–2482. https://doi.org/10.1021/acsphotonics.2c00586.

(42) Fieramosca, A.; De Marco, L.; Passoni, M.; Polimeno, L.; Rizzo, A.; Rosa, B. L. T.; Cruciani, G.; Dominici, L.; De Giorgi, M.; Gigli, G.; Andreani, L. C.; Gerace, D.; Ballarini, D.; Sanvitto, D. Tunable Out-of-Plane Excitons in 2D Single-Crystal Perovskites. *ACS Photonics* **2018**, *5* (10), 4179–4185. https://doi.org/10.1021/acsphotonics.8b00984.

(43) Fieramosca, A.; Polimeno, L.; Ardizzone, V.; De Marco, L.; Pugliese, M.; Maiorano, V.; De Giorgi, M.; Dominici, L.; Gigli, G.; Gerace, D.; Ballarini, D.; Sanvitto, D. Two-Dimensional Hybrid Perovskites Sustaining Strong Polariton Interactions at Room Temperature. *Sci. Adv.* **2019**, *5* (5), eaav9967. https://doi.org/10.1126/sciadv.aav9967.

(44) Anantharaman, S. B.; Stevens, C. E.; Lynch, J.; Song, B.; Hou, J.; Zhang, H.; Jo, K.; Kumar, P.; Blancon, J.-C.; Mohite, A. D.; Hendrickson, J. R.; Jariwala, D. Self-Hybridized Polaritonic Emission from Layered Perovskites. *Nano Lett.* **2021**, *21* (14), 6245–6252. https://doi.org/10.1021/acs.nanolett.1c02058.

(45) Anantharaman, S. B.; Lynch, J.; Stevens, C. E.; Munley, C.; Li, C.; Hou, J.; Zhang, H.; Torma, A.; Darlington, T.; Coen, F.; Li, K.; Majumdar, A.; Schuck, P. J.; Mohite, A.; Harutyunyan, H.; Hendrickson, J. R.; Jariwala, D. Dynamics of Self-Hybridized Exciton–Polaritons in 2D Halide Perovskites. *Light Sci. Appl.* **2024**, *13* (1), 1. https://doi.org/10.1038/s41377-023-01334-9.

(46) Hansen, K. R.; Wong, C. Y.; McClure, C. E.; Romrell, B.; Flannery, L.; Powell, D.; Garden, K.; Berzansky, A.; Eggleston, M.; King, D. J.; Shirley, C. M.; Beard, M. C.; Nie, W.; Schleife, A.; Colton, J. S.; Whittaker-Brooks, L. Mechanistic Origins of Excitonic Properties in 2D Perovskites: Implications for Exciton Engineering. *Matter* **2023**, *6* (10), 3463–3482. https://doi.org/10.1016/j.matt.2023.07.004.

(47) Song, B.; Hou, J.; Wang, H.; Sidhik, S.; Miao, J.; Gu, H.; Zhang, H.; Liu, S.; Fakhraai, Z.; Even, J.; Blancon, J.-C.; Mohite, A. D.; Jariwala, D. Determination of Dielectric Functions and Exciton Oscillator Strength of Two-Dimensional Hybrid Perovskites. *ACS Mater. Lett.* **2021**, *3* (1), 148–159. https://doi.org/10.1021/acsmaterialslett.0c00505.